# Hype in Science Communication:
## Exploring Scientists' Attitudes and Practices in Quantum Physics

María T. Soto-Sanfiel[1,2], Chin-Wen Chong[1], José I. Latorre[3,4]

1. Department of Communications and New Media, National University of Singapore, Singapore.
2. Centre for Trusted Internet and Community, National University of Singapore, Singapore.
3. Centre for Quantum Technologies, National University of Singapore, Singapore.
4. Quantum Research Centre, Technology Innovation Institute, United Arab. Emirates.

## Abstract

An interpretive phenomenological approach is adopted to investigate scientists' attitudes and practices related to hype in science communication. Twenty-four active quantum physicists participated in 5 focus groups. Through a semi-structured questionnaire, their use of hype, attitudes, behaviours, and perspectives on hype in science communication were observed. The main results show that scientists primarily attribute hype generation to themselves, major corporations, and marketing departments. They see hype as crucial for research funding and use it strategically, despite concerns. Scientists view hype as coercive, compromising their work's integrity, leading to mostly negative feelings about it, except for collaborator-generated hype. A dissonance exists between scientists' involvement in hype, their opinions, and the negative emotions it triggers. They manage this by attributing responsibility to the academic system, downplaying their practices. This reveals hype in science communication as a calculated, persuasive tactic by academic stakeholders, aligning with a neoliberal view of science. Implications extend to science communication, media studies, regulation, and academia.

**Introduction**

In recent times, science communication has garnered increased attention within the field of media studies (Bucher, 2020), paralleling the growing demand for a science of science communication aimed at facilitating meaningful dialogues between scientists and the public (Hilgard & Nan, 2017). Science communication, broadly defined, encompasses communication acts related to scientific work and its outcomes (Bucchi & Trench, 2014). It serves to relay scientific knowledge, address ethical and societal aspects, and facilitate interactions between scientists and various audiences (Kahan et al., 2017; Trench & Bucchi, 2010).

Science communication is multifaceted, characterized by diverse formats, channels, stakeholders, and often conflicting objectives (Marcinkowski & Kohring, 2014). It distinguishes itself from political discourse by upholding rigorous norms, including the comprehensive consideration of evidence, precise articulation of certainty levels, and the exact definition of analyzed phenomena (Jamieson, 2017).

The significance of science communication has surged in recent decades, accentuated by the need to address critical scientific challenges and rapid breakthroughs, which carry profound societal implications (Caufield, 2018). Effective science communication is considered paramount in the contemporary landscape, given science's growing role in national economies and increased competition for innovation advantages (Weingart et al., 2021).

Science communication, as highlighted by Rose et al. (2020), is a crucial element for building trust in science and supporting informed decision-making. This is especially significant due to the intricate interplay of factors like organizational reliance on funding, public financial support for science, state accountability, new public management practices, and the imperative of transparency (Marcinkowski & Kohring, 2014). The transformation of science communication over the years, starting in the 1980s, involves a shift from one-way information dissemination to fostering active engagement and dialogue across diverse audiences (Schäfer, 2009). This transformation has been further accelerated by the rise of social media (Stilgoe et al., 2014). In the current landscape, effective science communication is considered essential for enhancing knowledge production, workforce development, and promoting innovative capacity, facilitating informed decisions about scientific and technological programs (Weingart et al., 2021).

In this context, the concept of medialization, as explained by Weingart (2012) and Bucher (2020), is fundamental. Medialization represents the evolving and deepening connection between media, science, and the public sphere, encompassing expanded media coverage of scientific topics, specialized formats, and the involvement of new actors like Public Relation (PR) professionals and YouTubers. It also involves the adaptation of science communication to align with media standards.

As described by Marcinkowski & Kohring (2014), science communicators encompass various roles, from individual academics presenting their research in diverse settings to institutional communicators affiliated with academic institutions. These institutional communicators are responsible for disseminating the findings of their associated researchers to various audiences, especially general audiences or media communicators. External observers, often journalists or media experts, convey research processes and outcomes within a societal context

**Scientists' involvement with science communication**

Recent research has illuminated the evolving landscape of science communication and the role of academic scientists, acknowledging the centrality of science in modern society (Rose et al., 2020). Academic scientists now find themselves entrusted with the vital responsibility of effectively conveying scientific knowledge to diverse publics. This shift marks a significant departure from the traditional view of science communication as a peripheral duty, highlighting the contemporary cultural shift in academia towards greater public engagement. In this changing landscape, academic and research institutions are increasingly emphasizing media engagement and outreach to the public, the media, and social platforms (Rose et al., 2020).

Simultaneously, scholars face mounting pressures to capture the attention and approval of various stakeholders, including reviewers, editors, readers, funders, and promotion boards. This pressure stems from the competitive academic environment characterized by a growing number of scholars, journals, published papers, publishers, and authors writing in non-native languages (Hyland & Feng, 2021). As a result, scientists, much like journalists, are now in competition for visibility.

On the other hand, the increasing emphasis on productivity metrics in academic progress has heightened the pressure on scientists to highlight their findings and emphasize remarkable results to capture interest. This competition for attention has shifted the concept of "public engagement" towards marketing and PR, potentially leading to conflicts between faculty and management and affecting trust in science (Weingart, 2022).

**Hyping science**

Given the frequent need of researchers to promptly capture attention, engage with policymakers, and secure research funding, they might be prone to exaggerate their work's significance (Brown, 2003; Bubela, 2006). Academic pressure might compel them to employ hype into scientific discourse, rendering it a pervasive element of communication (Caulfield & Condit, 2012; Nerlich & McLeod, 2016).

In science communication, hype is defined as the use of promotional, hyperbolic, and dramatic language to glamorize, embellish, or exaggerate aspects of science, research, or scientists (Millar et al., 2020; Jones, 2017). It involves simplifying and sensationalizing science to shape future visions and gain support, often leading to optimistic or pessimistic expectations influenced by promotional activities (Roberson, 2020). Some authors view hype as a form of scientific deception characterized by misleading or unjustified communication (Wilson, 2019; Intemann, 2022).

Science hype is a complex phenomenon influenced by multiple actors, including scientists, media channels, and politicians (Caulfield & Condit, 2012; Jiang & Qiu, 2022). Media outlets are often criticized for sensationalizing science and prioritizing less rigorous research (Sumner et al., 2014). Some scholars attribute shared responsibility to scientists, press officers, business figures, and traditional media (Marcinkowski & Kohring, 2014). Recent studies highlight the active role of the public in propagating hype, especially in the context of social media-dominated information ecosystems (Taschner et al., 2021). Improving the quality of science-related news is crucial, with the academic community playing a vital role (Sumner et al., 2014). Additionally, Lerchenmueller et al. (2019) found gender-related disparities in framing research findings in scientific paper abstracts and titles, with male authors often using more positive language, particularly in high-impact journals, which is associated with higher citation rates.

## Consequences of hype

Authors have raised concerns about the adverse consequences of hype in scientific communication. Auch (2018) and Powers (2012) argue that creating and promoting expectations through 'hype' may erode public trust in scientific work, potentially leading to the spread of biased beliefs or political agendas. Saitz & Schwitzer (2020) state that sensationalized portrayals of science may foster confusion and unrealistic expectations, undermining public trust. Hopf et al. (2020) suggest that scientists endorsing 'hype' risk damaging the field of science and their own credibility, potentially leading to the spread of pseudo-scientific information. Millar et al. (2019) contend that prioritizing the marketing of research over its genuine significance may hinder impartial assessments of new knowledge. Some authors view 'hype' as a distraction that can overshadow other scientific topics, diminishing the effectiveness of scientific messaging (Roberson, 2019). Tiffany et al. (2022) indicates that the medialization turns scientists into instruments of media exposure and impacts scientific integrity and autonomy, potentially harming scientists' careers. Lastly, hype would stir emotions of envy and fear, emphasizing the multifaceted nature of 'hype' in scientific communication and the potential risks it poses to the integrity of science and public trust (Ezratty, 2022).

On the contrary, Roberson (2019, 2020) considers 'hype' to be an effective communication tool that drives engagement in the domains of science and technology, promoting progress and advancing discussions on feasibility and potential. This perspective suggests that the advanced rhetoric used in scientific discourse propels competition, global leadership, and societal benefits (Roberson, 2019). Caulfield and Condit (2012) support this view, arguing that empirical data does not consistently show that hype is inherently negative, as audiences are not always influenced by it, leading to creative debates on shaping social futures.

## Objectives and Research Questions (RQs)

This study explores scientists' perceptions and use of scientific hype in media, offering insights into how they introduce and present their breakthroughs and their concerns about public engagement (Caufield, 2018; Tiffany et al., 2022). It responds to the need to assess how scientists communicate their research in the media and its impact on public and policy discussions (Caulfield et al., 2021; Kousha & Thelwall, 2020) while examining their attitudes in science communication, particularly regarding hype (Rose et al., 2020; Powers, 2012).

This study addresses the following research questions:

RQ1: How do scientists employ hype in their research communication?

RQ2: What are scientists' attitudes and perspectives on the use of hype in science communication?

RQ2.1: How do scientists' beliefs, emotions, and behaviors influence their use of 'hype' in science communication?

RQ2.2: What are scientists' views on the origins, sources, reasons, and consequences of 'hype' in science communication?

The phenomenon of hype in science communication has garnered increasing research attention due to its widespread presence across various communication formats, including articles, abstracts, and grant applications. While studies have explored the consequences of hype, there remains a significant gap in our understanding of scientists' attitudes and practices regarding this phenomenon. Scarce research reveal that scientists often find themselves at the center of hyping science, driven by the fierce competition for research funding

Chubb & Watermeyer (2017) observed academics from Australian and British universities and revealed that the pressure to secure research funding motivates them to employ sensationalist strategies, normalizing the use of exaggerated impact claims as a means of thriving in academia. Some even describe themselves as "impact merchants," striving to meet performance expectations set by senior managers. This research portrays academics as adept storytellers who craft persuasive narratives about uncertain futures while also acknowledging their complicity in the system they critique.

A study by Miller et al. (2020) delved into the use of hyped language by first authors of medical papers associated with a Japanese medical school. These scientists employed hype to instil confidence in their research methods, emphasize precision and thoroughness, underscore the novelty of their study components, and persuade readers about the clinical significance of their work. This persuasive language was directed at convincing policymakers and clinicians to embrace new evidence. The study concluded that effective medical scientific writing involves striking a balance between objective truth and subjective interpretation, making it a form of art.

While media "hype" has received some research attention (van Atteveldt et al., 2018), it has been explored less as a distinct concept in communication and media studies (Powers, 2012). Existing research often focuses on how "hype" spreads, emphasizing media coverage intensity. For instance, Vasterman (2004, 2005, 2018) views media "hype" as a significant news wave driven by specific events, amplified through news production processes. Another perspective defines "hype" as a genre related to promotion, marketing, and advertising, with Gray (2008, 2010a, 2010b) introducing the concept of paratextuals shaping "hype" meaning.

Our research innovatively examines scientists' attitudes towards hype, covering their thoughts, emotions, and behaviors. Departing from the traditional view of hype as merely a communicative act (Power, 2012), this approach addresses a significant research gap by offering insights from the scientist's perspective, as exemplified in works such as Nisbet et al. (2015).

## Method

This research employs an Interpretative Phenomenological Approach (IPA) known for its effectiveness in exploring individuals' experiences and providing insights into social contexts and power structures (Finlay, 2011; Pietkiewicz & Smith, 2014). IPA enables a nuanced understanding of how participants interpret their experiences, reflect on their actions, and make sense of their emotions (Neubauer et al., 2019). The use of open and exploratory research questions, as in this study, encourages individuals to delve into their thoughts and behaviors (Miller et al., 2018), facilitating a comprehensive analysis of their lived experiences and the meanings they derive (Chan & Farmer, 2017).

**Participants and procedure**

To understand scientists' attitudes toward 'hype,' we conducted online focus groups (FGs) using Zoom, a platform suitable for remote discussions (Kite & Phongsavan, 2017) and known to yield data comparable to in-person FGs (Keemink et al., 2022). This approach aligns with IPA as a data collection method (Love et al., 2020).

Research literature suggests that four FGs are sufficient for exploring various aspects of a phenomenon (Guest et al., 2017; Hennink et al., 2019). The first group identifies key topics, while the subsequent two delve deeper. Additional groups would not significantly enhance insights (Hennink et al., 2019). Despite reaching data saturation after the third FG, we conducted a total of five online FGs to ensure a comprehensive and valuable dataset. These online FGs took place between October and November 2022.

The study included 24 Quantum Physicists. Of these, 8.33% identified as women, and 91.67% as men. They were active researchers at universities and international research institutes in the United Arab Emirates, representing diverse nationalities and educational backgrounds from around the world. These participants spanned various career stages, from post-doctoral students to senior researchers, providing a broad range of perspectives. Recruitment utilized a snowballing technique, starting with the researchers' social networks and email outreach. Participants volunteered and did not receive compensation, and signed informed consent was collected. Quantum physics is experiencing significant hype, with overstated claims about quantum computing's potential (Ezratty, 2022).

The online focus group discussions, with 4-6 participants each, followed a semi-structured questionnaire format. Questions included whether they had engaged in professional hype, reasons for doing so, causes of hype, actors responsible for it, its effects, potential victims, and the most hyped topics in their field. We provided a comprehensive definition of hype to capture various attitudes and experiences related to it, clarifying that it involves simplifying, exaggerating, and sensationalizing scientific information at all research and communication stages. An unfamiliar researcher led the focus group, while another unfamiliar researcher served as an assistant and observer.

The interviews were audio-recorded, and verbatim transcripts were created from these recordings.

The procedure was ethically approved by the National University of Singapore.

**Analysis**

We conducted an inductive thematic analysis (TA) using Atlas.ti software to analyze the transcriptions. TA, a widely-used qualitative data analysis method in the Social Sciences (Guest et al., 2012), systematically reviews textual data to identify key themes and interpret their underlying structure and content (Stebbins, 2001).

TA aligns with the existential perspective of IPA (Willig, 2007), which aims to illuminate nuanced meanings and individual perspectives rooted in personal experiences and understandings. As TA, IPA values individual interpretations as unique and valuable narratives, shedding light on various aspects of participants' experiences and their grasp of the subject matter, enriching the understanding of the phenomenon under study. The integrated approach of TA and IPA captures the complexity of the explored phenomenon (Spiers & Riley, 2019).

Accordingly, our methodology encompassed multiple steps, including identifying significant themes and facets within participants' responses, exploring rich narratives, uncovering underlying meanings and emotional nuances tied to these themes, and comparing them across different understandings. Individual experiences were leveraged to illustrate these thematic aspects. The research data is available upon request.

## Results

### Who hypes science?

When asked about the agents of hype, participants offer a variety of perspectives. However, participants primarily point to scientists themselves as the primary generators of hype. Next, we provide detailed descriptions of the scientists associated with the hype.

**Scientists recognize they employ hype.** The consensus among most participants is that scientists indeed embrace hype. One participant shed light on why scientists resort to this practice, attributing it to the increasing difficulty of securing funding, as he explained, "this is a good way to get funding. To go in this wave of hype and try to get funding from that, which typically most of researchers then use these to do nice, scientific investigations." These sentiments resonated with the majority of participants, who also expressed concerns about scientists employing hype as a funding acquisition strategy.

**Major corporations are also responsible for hype.** Conversely, a similar proportion of participants argue that significant hype in science communication can be attributed to major corporations. One participant pointed out that hype can be linked to the active involvement of renowned companies like Microsoft, IBM, and Google in the specific field of quantum physics, noting: "These companies used to really push the entire field, which was a very good thing. I mean, the reason why we have so much money [is] because these companies very professionally like this research and also achieved milestones together with a huge, huge PR effect." Another participant concurred regarding the substantial investments made by these major companies in quantum physics. Finally, another participant argued that there was agreement in the research community that Google's contribution to the hype in quantum physics was notably significant, particularly due to their supremacy experiment (see Gibney, 2019).

**Marketing and PR departments of research institutions often use misleading or inappropriate hype.** Some participants attributed the hype in science communication to the marketing departments within research institutions. As one participant noted, "the usual hype is when you go to the marketing department and say what you did and they make out of it … some cool story, some big story." Overall, the participants had a negative attitude towards the research centers' marketing departments due to their hype tactics, leading to feelings of frustration, anxiety, and anger.

**Startups are often the culprits in hype practices.** A minority of participants identified startups as the responsible parties in the practice of hype, driven by their imperative need to secure funding. One participant succinctly remarked: "now the hype is generated by smaller companies."

**Journalists and media must be blamed for hype.** A subset of participants shared the perspective that hype often originates from journalists and the media. One participant observed that journalists tend to align their stories with their audience's interests, drawing from their limited understanding of the subject matter, occasionally resulting in stories that are more distorted or exaggerated than necessary. Another participant emphasized the disparities between the hype generated by scientists and that by journalists. According to this participant, "the hype produced by us and the journalists is different…hype produced by the media can be somewhat distorted from our hype. Only because journalists don't have a scientific background."

**Privileged and powerful individuals, mostly men, hype.** Another few participants believe that hype is often generated by individuals in positions of power, specifically privileged people and specific countries with substantial wealth. A female participant expressed the view that someone seeking funding and the opportunity to conduct research typically needs to possess charisma, an extensive network, and the time and resources necessary for deep concentration. She noted that "hype is driven by the ego of white, privileged males in science." Another female participant suggested that hype tends to occur in places where there is less necessity to address other pressing issues, which can be attributed to "white privilege that drives the hype themselves and, in a dozen, other rich countries."

**Scientists from certain countries tend to hype more.** A few participants believe that different regions' funding systems impact the use of hype in science communication due to research type, investment levels, and participant characteristics. Specifically, one participant notes: "the American system is more practical, but the European system is more fundamental." Consequently, as stated by another participant European researchers, focused on basic science, have less need for hype, explaining: "In the US they are more focusing on applications, on things that can be marketed there." Another participant highlights the financial aspect, stating that the greater hype in the US is " part of the business model [which] give millions to somebody and hype it, in hopes that there's greater adoption." Another participant attributes much of the hype to the United States and its startups, which "produce so much pieces of news about the things that [they] are doing. So, most of the hypes... come from those startups."

**Pseudoscientists hype.** One participant contends that hype frequently emerges from the actions of pseudoscientists, whom he categorizes as individuals "who don't really know what they're talking about". However, he also concedes that among these individuals, there may be those who possess knowledge about the topics they address, classifying them as "dishonest people."

## What drives the occurrence of hype?

Participants were asked about the reasons for the existence of hype in the communication of science. In response, they highlighted that hype in science communication serves a dual purpose. It helps secure funding for research and reflects the commercialization of academia. On the other hand, that major entities, including corporations and countries, employ hype to demonstrate dominance, with scientists being instrumental in this power play. Next, we provide a comprehensive description of the reasons identified by scientists that contribute to hype.

**Obtaining funding for scientific research.** When asked about the reasons for hype, most participants acknowledge its pivotal role obtaining funding for scientific research and product development, as introduced earlier.

**Academic funding system.** However, most of the participants view hype as a problematic practice rooted in broader organizational issues within the academic funding system. This is exemplified by one participant's observation stating that "the whole funding system of the academy…kind of forces you into twisting a little bit your story to attract funding". Today, scientists are esteemed for their ability to secure funding, and this pressure often compels them to adapt and embellish their narratives. It's seen as an institutional challenge, not a reflection of scientists' intrinsic desire to hype their work.

**A profit-driven perspective on science with a commercial focus.** Accordingly, the majority of participants consider research today as an activity that must be sold to different entities and hype contributes to it. One participant said that "You need to sell. It's a skill and it's necessary to get money [for research]". Another participant agreed that he primarily has to sell his work to the journals to get his work published through making "a good abstract and introduction and conclusions where you try to sell yourself, to be honest".

**Thriving in the fiercely competitive world of modern science.** On the other hand, for some participants, the existence of hype in the academic world is closely tied to their survival within it. As one participant succinctly puts it: "A professor at the university doesn't have a choice if he wants to grant, he has to jump on the hype train." This sentiment is further reinforced as another participant shares anecdotes about colleagues and friends who feel compelled to produce attention-grabbing papers in prestigious journals like Nature to remain competitive in their academic careers. He notes that this often necessitates a level of "hypeness," where individuals may feel they need to exaggerate their work to secure their place in the academic landscape. Another participant concurs, highlighting that, in the current scientific scene characterized by an overwhelming volume of research articles, particularly young scientists must actively promote themselves to stand out. According to participants, academia values people with more publications.

Conversely, the older, more senior participants in the sample strongly reject and critique hype in scientific communication, attributing it to contemporary times. In contrast, younger participants acknowledge the system's existence and, while disapproving of excessive or distorted hype, tend to take a more pragmatic approach. The former group generally has well-established careers, while many in the latter group are in the early stages or have recently solidified their careers.

**Rivalry between corporations and nations.** A subset of participants suggests that competition, both among companies and countries, also plays a pivotal role in generating hype. One participant emphasized that "this hype is being driven by companies that are competing, like Google and IBM." Another participant attributes the enthusiasm for hype among companies to the relative lack of regulation they face. He points out that "they're private companies with closed capital. So they are not under the jurisdiction of the [United States] Securities and Exchange Commission. So they can make the wildest PowerPoints they can do." Another participant adds that sensationalist articles and hype produced by local industries can instil a sense of national or communal pride, making people feel as though they possess a cutting-edge technology or skill set that sets them apart from others, following the trail set by interests imposed by private companies pursuing their specific commercial goals.

**Shaping expectations regarding scientific outcomes.** The deceptive nature of hype becomes apparent in the following comments from two participants. One individual suggests that some scientists may resort to hype as a means to buy time, implying a conscious manipulation of information and expectations. He explains: "We are just kind of exaggerating a bit. Maybe we believe it will be ready in ten years, but then we say, okay, maybe next year, you know, and then we try to get from there next year, we will try to convince you that one more year will be enough." This observation aligns with another participant who acknowledges that hype's existence can be attributed to pseudoscientists. This participant distinguishes between those who unknowingly make scientifically unsubstantiated claims and those who intentionally engage in deceptive practices for dishonest profit.

## When and where does hype occur?

Participants were asked where hype is found and if that appearance correlates with a specific moment of the investigation process. They identify a diverse range of contexts where hyped science communication can occur, which are described next.

**During the submission and peer-review process of journal articles.** Some participants pointed out that hype is prevalent in the submission and selection of journal articles, attributing responsibility to both researchers and editors for its prevalence. For instance, hype tends to be common in the abstracts, summaries of articles or cover letters submitted to journals by researchers. One participant explained: "if you want to publish your article in the journal, sometimes they [journal's editors] ask for a cover letter or a short summary of what you're doing. And you're actually quite encouraged to hype up your own research." Another participant emphasized that journal editors often take notice of specific popular phrases, such as "quantum advantage" or "quantum computers," when making decisions about article acceptance.

**In promotional materials created by research centres.** On the contrary, a few participants believe that hype is prevalent in the marketing materials produced by research centres. One participant shared their experience, explaining his experience with an organisation with very aggressive PR: "so initially I tried to challenge it, to change it, and it didn't work out. So I changed my strategy. So I agreed to PR articles to be hyped, but I checked it very much on the wording…. if you check it sentence by sentence, there will be not a single false sentence." This example illustrates that scientists often do not have the final say in the decision to employ hype. Instead, it can sometimes be driven by the marketing teams of their research centers. This perspective is reinforced by another participant's experience, who mentioned that when he sought to publicize a breakthrough achieved by his team, the marketing team wanted to hype the press release. He noted: "To this day, they have not published it because I refused to allow them to hype it. They want to hype."

**In conferences.** One female participant observed the presence of hype at a conference featuring a quantum booth. In her account, this participant described her research center's involvement in the conference, where scientists interacted with attendees. She explained: "I had to be a bit more optimistic about quantum computing than I usually am because I was kind of selling. We don't have a proper product to sell, but I was selling the company, so I couldn't say it won't work. So, it was kind [of] engaging in the hype."

## Personal experience with hype

Participants were asked if they had incurred any form of hype during their career and why. In general, their responses showed they strategically wield the tool of hype, with their attitudes shaped by their audience. Next, we include detailed characterizations of scientists' personal experiences with hype.

**I have enhanced abstracts and scientific article content with hype.** As advanced earlier, most participants admitted to intentionally introducing an element of hype in the abstracts and within the text of the articles submitted to journals for publication. One participant emphasized the necessity of this practice, stating: "I learned that you need some hype to get things pushed, so you need them. In your introduction of those scientific publications, you need to mention that there's a lot of interest on that, that the field is pushing towards that or else your paper will never get accepted.'"

**I have hyped in applications for research funding and grant proposals.** One participant acknowledged incorporating "some kind of hype" into their research proposals and grant applications, explaining it as "upselling what you do so that people invest and they give you money to continue doing it." Another participant expanded on this, noting that "many proposals require you to predict in some sense what your research will bring to society," and stressed the need to align with prevailing hype by saying "what everybody else says," as failure to do so would result in a lack of funding.

**I have hyped in corporate events and conferences.** A few participants confessed to engaging in hype when they took part in a conference. One participant noted: "I think we participated in the hype just by the fact that a quantum stand existed in a corporate event." Another expressed the importance of participating in such events, stating "I think we should participate in more of this kind of events because it's important. It's an opportunity for us to talk to people and to understand people, the general public's perception of quantum physics in our case, and it's an opportunity to talk to them and to make some of them understand what's actually going on and what is overhyped about it."

**I have hyped when engaging with individuals outside of their field.** A few participants openly admit to intentionally injecting a level of optimism or 'hype' into their discussions when engaging with individuals outside the field, as opposed to their interactions with colleagues within it. One participant articulated this contrast, stating "when we talk among us, we are much more pessimistic about the results and the future of the field than we are ourselves when we talk to people outside… when we're talking among us, we're like they are trying to sell this idea; this is obviously not going to work that way or it's not going to be ready in five years, but in 20 years.' But then when we go outside our environment and then people ask us about what we think about quantum computation, and then we start being more optimistic." Another female participant added to this, noting: "Usually we say it's guaranteed… It's not 'there is a possibility that we will have those algorithms work', it's 'no, it will happen'"

**I have navigated a wave of hype.** A few participants acknowledged their engagement with 'hype' in research, with a sense of fortunate alignment. One participant expressed this feeling, stating: "I was lucky to be on a couple of hypes before. I didn't choose them, it was just how the career path led me, life led me to those parts. I did my Masters in solar cells and back then in 2014, there was a lot of funding for that. There's a lot of money being pumped into that." Another participant, however, shared a contrasting perspective, feeling lucky in their research against the prevailing hype.

**I have hyped when disseminating my research through social media.** One participant believes that he participates in hype through his use of social media to discuss quantum physics. He elaborated, saying: "I now have a Twitter for Quantum, because it makes my research more visible. So there, you also tweet, like, 'look at this new result, this is awesome.'" Another participant actively participated in hype by producing content for his marketing department, specifically tailored to social media. He mentioned the collaboration between his marketing department and YouTuber Nash Daily, stating: "If you see any of his videos, it's all full of exaggeration…. It's the way that YouTubers communicate.'"

## Consequences of hype

When queried regarding the ramifications of hype in science communication, the majority of participants predominantly cited adverse consequences, while only a minority discussed potential positive outcomes stemming from hype. Next, we provide a detailed account of the participants' views on the consequences of hype.

**The risk of funding withdrawal due to unmet promises.** Several participants highlight that one of the significant consequences of hype is the potential withdrawal of funding if promised results cannot be delivered. One participant pointed out the predicament scientists face when they fail to meet the high expectations set by their own hype, stating: "what happens in two or three years when we are supposed to give some products or yield some results, the research, and we don't have anything and we might be quite sure that we won't have anything of the sort that we have promised… we will lose the fund, there is this sort of problem like we've promised too much." This sentiment was echoed by another participant who emphasized the connection between excessive promises and a shortage of tangible outcomes, leading policymakers to contemplate funding reductions. He explained: "if there's too much overpromise and too little deliverables, things which are really shown and work, then at some point policymakers will start saying, 'okay, we have to cut the money.'"

**Research field stagnation.** Additionally, one participant expressed concerns about the potential impact on the field if the promises and hype are not realized, possibly leading to stagnation. This participant agreed that a decrease in funding could have adverse consequences for the field, as it would hinder the influx of new talent and resources required for its development.

**Damage to the scientific institution's reputation.** Several participants express concerns about the potential harm to the institutions' reputation when excessive hype is involved. One participant highlights the risks of hype in science communication, emphasizing the potential negative impact on how other institutions perceive the works of others. He articulates the concern, stating: "If we exaggerate in the way that we communicate, then other institutions are not going to perceive us as a serious scientific institution, and that is going to prevent us from having the best talent or attracting talent." This concern is echoed by another participant: "my main concern is that if we have this image as an organization to produce untrue statements, you scare away good scientists. They wouldn't come anymore and this would harm us..."

**Research concentrated on areas with significant hype.** Several participants express concerns about the potential consequence of hype on research focus, particularly in favour of a few specific areas. One participant points out that hype can lead to a concentration of funds within a narrow segment of the scientific community, leaving other scientists who are working on different areas with fewer articles and less funding. They caution: "hype might also funnel funds into a very small part of the community that is doing the hyped-up thing and that might leave the wider community of scientists that might be focusing on some other things, that have not as many articles as the other one, to be left without funding at all." Another participant adds to this perspective, highlighting that hype tends to elevate one narrative or idea to the forefront, drawing significant attention and subsequently steering the scientific community in that direction.

**Eroded trust in science.** Some participants express concerns regarding the potential negative impact of bad hype on the public's trust in science. One participant emphasizes that scientists aim to provide the public with information to build trust between the scientific community and the general audience. They highlight the importance of effective communication to maintain this trust and express the consequences of poorly communicated or misguided hype: "scientists give this public information to build the trust between the scientific society and a large audience so that people understand where this money are [is] going … if it's going wrongly, or badly communicated, then people are [will] losing [lose] this perception and say why not spend this money on better roads or whatnot."

**Misleading stakeholders into believing in certain achievements' feasibility.** One participant identifies a significant adverse outcome of hype, which is the potential for misleading stakeholders into believing that certain achievements are feasible when, in reality, they may not be possible, at least not in the short term. He elaborates on this concern, stating: "it [hype] has a negative impact when the exaggeration misleads different stakeholders into believing that things are possible, which are not possible or not, at least not in the short term. And that can lead later on to disappointment…" The same participant also raises the issue of distinguishing reality from exaggeration, emphasizing the challenge derived from people who cannot discern between genuine advancements and exaggerated claims.

**Rising of science deniers and those who see science as unattainable**. One participant highlights a detrimental outcome of hype, particularly when it leads to people failing to comprehend a scientific topic and ultimately denying its authenticity. He expresses this concern, stating: "But I feel that some articles which convey the message in such a way that make people get away from science and feel like it's something 'I can never understand. It's something I cannot grasp in any way'. And I think this is dangerous." This participant further explains that such depictions of science can contribute to the emergence of science deniers and individuals who perceive science as an unattainable realm. He elucidates that these people may believe that "science is something so far away that I cannot even try to understand it." The participant underscores the importance of a well-informed society for promoting fairness and societal betterment, emphasizing that people need access to information about science.

**Emergence of pseudoscience.** One participant raises concerns about the potential emergence of pseudoscience as a consequence of hype. He illustrates this point by highlighting how, when people attempt to promote scientific concepts and include details that "might not be very close to reality," hype can lead to the propagation of pseudoscientific ideas. The participant provides examples such as "ideas … like the 'quantum healing,' … 'quantum mattresses,' and … other quantum-related concepts that are pseudoscience." Additionally, the participant emphasizes the significance of being cautious and responsible in how science is communicated. He suggests that if individuals are not careful and responsible in their communication of science, it may inadvertently create fertile ground for the spread of pseudoscientific beliefs.

**A positive outcome: sparking public interest and attracting investments.** A minority of participants recognized positive outcomes of hype in science communication, highlighting how it can stimulate public interest and attract investments. One participant expressed this view by stating: "it's nice to have this sort of hype in a way because it generates, promotes attention to the field." Another participant concurred, emphasizing that "with hype you managed to attract maybe not scientists, but the wider community into becoming interested about a topic and getting to know more. And also, that has some potential good consequences in that it attracts investors and investors can fund research if they consider that interesting and useful."

**Are there victims of hype?**

Participants were queried about their perception of potential victims of hype and who these victims might be. In general, they state that hype can have various victims. Their perspectives are summarized next.

 **Scientists' careers.** Some participants expressed the view that scientists can become entangled in the web of hype and be victims of hype. One participant conveyed the impact of hype on academic careers, suggesting that it can lead to significant shifts, remarking: "individual careers kind of break because they just missed out on the hype that's in the field." Another participant shared similar concerns, emphasizing that prolonged failure to meet the heightened expectations could result in dwindling investments. However, there was a counterpoint offered by one participant who raised the question of whether scientists could genuinely be considered victims of hype, pondering: "if you are part of the creation [of hype] process, then I don't know if you can call yourself a victim."

 **People with limited scientific understanding.** Several participants expressed the belief that individuals who lack a deep understanding of science are victims of hype. A female participant recounted encounters with people who lack comprehension of the work done by scientists, and as a result, harbor fear and apprehension toward sciences. She highlighted the common sentiment that quantum physics is potentially perilous, drawing parallels to past concerns about scientific advancements due to hype in media. For instance, during the construction of the Large Hadron Collider at CERN, there were unfounded fears of a black hole appearing when the device was activated due to negative press coverage. A consensus emerged among the participants that misinformation and fear, often fueled by a lack of understanding and hyped press coverage, are central issues at the heart of these concerns. One participant concurred, emphasizing, "the main issue is probably misinformation or fear."

 **Investors.** One participant exhibited an awareness of the possible manipulation of investors by scientists who employ hype as a strategy, leading to adverse consequences for these investors. The participant emphasized the potential harm that could be inflicted, stating: "if you're hyping to gain money from an investor, you might be causing damage to a specific person or group or company."

 **Degree-seeking students.** One participant expressed concerns about the potential impact of hype on students pursuing degree programs, particularly those highly specialized in specific topics like quantum physics. He noted: "You create hype, it creates a lot of investment, a lot of money, a lot of people interested in the topic. And then if the hype goes down, and the investment goes down, what happens to those people?" The participant also emphasized the challenges students may face when their educational focus becomes highly specialized, potentially limiting their ability to adapt or change career trajectories if the field experiences a downturn.

 **Decision-makers shaping the research and funding landscape.** Finally, one participant raises concerns about the potential victims of hype, specifically individuals in influential positions within the research and funding landscape. He mentioned that people like "the principal investigators who directly decide which projects we work on, to the funding level with the research councils, where they put out calls for proposals," would be among those most susceptible to the influence of hype.

## Discussion

In the pursuit of fostering meaningful dialogue between scientists and the public (Hilgard & Nan, 2017), and recognizing the multifaceted communicative strategies employed by science and scientists in disseminating their findings (Bucchi & Trench 2014), this research delves into scientists' interaction with hype. Diverging from the traditional focus on observing hype's dissemination in media communication studies, whether as a marketing genre (e.g., Gray, 2008, 2010a, 2010b) or its propagation (e.g., van Atteveldt et al., 2018; Vasterman, 2005, 2018), this research offers a fresh perspective. It examines hype as a manifestation of scientists' attitudes (Powers, 2012). This contribution is particularly valuable, considering the limited body of literature that explores science communication from the scientist's standpoint (e.g., Besley et al., 2018; Nisbet & Markowitz, 2015). This perspective is also essential because scientists fulfil a dual role as both initiators and recipients of hype within the domain of science communication.

The importance of studying scientists' use of hype is grounded in the evolving landscape of science communication and the phenomenon of science medialization. As academia increasingly emphasizes public engagement (Rose et al., 2020) and recognizes effective science communication as a vital part of scientists' roles in society (Marcinkowski & Kohring, 2014; Tiffany et al., 2022), scientists are under growing pressure to capture the attention of various stakeholders (Hyland, 2023). In this competitive and metrics-driven environment, scientists, akin to journalists seeking attention, may amplify their findings or prioritize publishing exceptional results (Fire & Guestrin, 2019). They also conform to media standards when interacting with the press (Bucher, 2020). Consequently, hype, involving simplification, exaggeration, and sensationalization of science (Roberson, 2020), has become prevalent among scientists (Brown, 2003; Brown & Michael, 2003; Nerlich & McLeod, 2016). Despite its prevalence, empirical studies on scientists' attitudes toward hype, crucial for effective science communication and academic recognition, have been surprisingly scarce in the field of communication studies. Our qualitative research addresses this gap by exploring the thoughts, emotions, and behaviors of quantum physicists, a field experiencing hype (Ezratty, 2022). This research responds to the need to examine how scientists communicate their research to diverse audiences, influencing future discussions in the public and policy domains (Caufield et al., 2021; Kousha & Thelwall, 2020), and contributes to the limited literature on scientists' attitudes toward different aspects of hype (Chubb & Watermeyer, 2017; Miller et al., 2020).

This research affirms that scientists view science hype as a complex phenomenon influenced by various factors, in line with prior research (e.g., Caufield, 2018). Participants in this study concur that hype originates from diverse sources, including scientific entities, media outlets (Caulfield & Condit, 2012), politicians (Jiang & Qiu, 2022), and business leaders (Marcinkowski & Kohring, 2014). These findings align with existing notions in the scientific community that consider hype an inherent aspect of the scientific arena, shaping scientists' interactions with various audiences (Brown, 2003; Brown & Michael, 2003; Nerlich & McLeod, 2016).

However, this study empirically affirms that scientists actively acknowledge their central role in generating science hype, in agreement with limited prior research (e.g., Sumner et al., 2014). The results also highlight their ability to tailor science communication for diverse audiences, objectives, and visibility strategies within academic and organizational contexts. Additionally, participants in this study attribute some responsibility for science hype to major corporations and research institution marketing departments, as well as, to a lesser extent, startups, journalists, privileged individuals, and pseudoscientists in its creation and dissemination. These findings emphasize the complex array of factors influencing science communication hype (Caufield, 2018; Marcinkowski & Kohring, 2014) and its structural foundations.

The results of this research affirm that scientists consciously navigate and utilize hype as an intrinsic and dynamic element, driven by the competitive nature of the field. They employ hype strategically in various contexts, such as securing funding, enhancing visibility in promotional materials, abstracts, and articles, emphasizing impact in grant proposals, addressing initial misconceptions at events, shaping their image when interacting with non-experts, and leveraging social media for broader dissemination. These practices align with the attention economy in science, where scientists use hype to capture rapid attention by magnifying their work's significance, reflecting the growing emphasis on "public attention" within the academic system and their own careers. This transformation underscores the evolving role of scientists as mediators of media exposure, exemplifying the process of science mediatization (Hyland, 2023; Brown, 2003; Bubela, 2006; Caulfield & Condit, 2012; Marcinkowski & Kohring, 2014; Tiffany et al., 2022; Bucher, 2020).

This research has delved into scientists' motivations for generating hype in science communication, revealing diverse reasons closely tied to specific objectives. According ti the participants, major corporations use hype to pursue financial gains and validate substantial investments. Research institution marketing departments employ hype to enhance research significance and public appeal for potential investors. Startups leverage hype to attract crucial funding. Journalists resort to hype for commercial success, aligning with audience interests. Pseudoscientists exploit science's credibility to market their ideas, regardless of scientific validity. These motivations underscore scientists' alignment with a neoliberal, competitive, commercial, de-professionalized, and service-oriented academic system (Chubb & Watermeyer, 2017; Rose et al., 2020; Bucher, 2020). The study also suggests that scientists in countries with robust public academic systems, like Europe, would engage in less hype compared to their counterparts in the US, which places a stronger emphasis on competition, private funding, industry-sponsored research, and market-oriented approaches.

Furthermore, the outcomes of this research unveil a spectrum of emotions interwoven with the broader concept of hype and its multifaceted practices in science communication, a dimension not previously explored in depth by existing research. Among the participants, a complex mix of emotions surfaces, primarily stemming from the belief that scientists resort to hype as a means of securing funding. This emotional tapestry elicits a sense of concern as participants perceive hype as both a necessary tool in the competitive funding landscape and a potential source of distortion and exaggeration of scientific information, occasionally giving rise to feelings of frustration and scepticism.

Participants express negative attitudes towards research institution marketing departments, citing emotions like frustration, anxiety, and anger due to their perceived role in generating hype and the challenges in science communication. Frustration is particularly evident among those who see hype as a byproduct of broader organizational issues within the academic funding system, expressing dissatisfaction with the pressure to embellish narratives to secure support. Desolation arises from the need to sell their research, reflecting the profit-driven aspect of contemporary science. Compulsion is also prevalent, with some participants feeling obligated to engage in hype for survival in academia, given its competitive and coercive nature. The competitive aspect, contributing to hype as a contest for recognition, is viewed negatively by senior scientists who've observed the shift to neoliberal practices. Younger scholars, starting their careers, are more pragmatic and less emotionally charged about the necessity for hype in the research system, displaying less intense emotions compared to their senior counterparts.

Participants express negative emotions, including frustration, anger, and resentment, when encountering deceptive practices by pseudoscientists related to hype. These emotions highlight their concerns about the adverse impact on the scientific community and its integrity. Some participants, particularly women, also express frustration with the influence of privilege and wealth in driving hype, with privileged males and wealthy countries seen as primary culprits, generating resentment and irritation. This gendered perspective aligns with the findings of Lerchenmueller et al. (2019), which reported greater use of hype by male scientists, making it a unique contribution to the study of gendered scientists' relation to hype.

Scientists' emotional responses to hype are influenced by the perceived source and its role in the scientific community, as well as the scientists' personal objectives. They view hype resulting from significant investments by major corporations positively due to its role in boosting research funding. On the other hand, scientists tend to feel more positively towards hype generated by collaborators who support the field economically and negatively towards those they perceive as lacking legitimacy in science communication. Overall, hype is seen as a pervasive and coercive influence within the competitive academic landscape, pushing scientists to engage in practices that may compromise their work or autonomy. Some feel steered toward objectives other than genuine knowledge, with hype practices serving as tools for asserting dominance, making scientists instruments through which power dynamics are demonstrated.

This research reveals a dissonance between scientists' extensive involvement in hype across communication contexts and the negative emotions it triggers. Applying Festinger's theory of cognitive dissonance (1957) becomes relevant in understanding the discomfort arising from conflicting beliefs, attitudes, or behaviors. Recent studies have explored cognitive dissonance among scientists (Schrems & Upham, 2020) and within fake news in science communication, underscoring the need for a comprehensive examination of scholars' emotional states and negotiations (Taddicken & Wolff, 2020). Schrems and Upham's study on environmentally committed scientists (2020) unveiled strategies for resolving cognitive dissonance, such as suppressing inconsistencies, denying control, justifying through benefits, and employing distraction. In the current study, scientists primarily justify their dissonance between negative feelings toward hype and participation by compensation and denial of control, often attributing responsibility to the academic/scientific system. Some compare their hype to non-legitimate actors, such as pseudoscientists or marketing and PR departments. They employ strategies of forgetting, focusing on the hype generated by others. This paradoxical situation, where scientists use hype to gain attention while diverting attention from their own dissonant practices, warrants further exploration in the field of science communication. Future research, particularly on hype practices, should consider applying this theory, taking into account personal attributes, such as gender and career stage.

The research findings suggest that scientists strategically employ hype, adapting their attitudes according to their audience. They maintain a cautious and critical approach when interacting with peers while adopting a more optimistic and enthusiastic persona when addressing the general public. This dual communication style helps them safeguard their professional image within academia while garnering support and interest from a broader audience. Some scientists view hype as a double-edged sword, acknowledging its ability to capture initial attention but also recognizing the need to rectify any initial exaggerations or misconceptions. This suggests a multifaceted role, including elements of manipulation. Overall, the scientists in the sample demonstrate a conscious awareness of hype's role in their work, utilizing it strategically with an appreciation of its potential benefits and ethical considerations, reflecting a pragmatic and adaptable approach to this prevalent aspect of their field.

In summary, this research highlights participants' concerns regarding the impact of hype in science communication across science, the scientific community, and society, with a predominant focus on the potential negative consequences. Concerns include the risk of losing funding if promised results aren't realized, damaging the reputation of scientific institutions, shifting research focus driven by hype, eroding public trust, and misleading stakeholders. There's also recognition of the risk of pseudoscience proliferating and the emergence of science deniers due to exaggerated scientific concepts. However, a minority of participants acknowledge a positive aspect of hype, perceiving it as a means to ignite public interest and attract investments, serving as the rationale for scientists' pervasive use of hype in their professional engagement with science.

Finally, the research findings propose that "hype" in science communication can be defined as a purposeful and persuasive attitude involving thoughts, emotions, and behaviors, which is adopted by various stakeholders (e.g., scientists, research institutions, funding agencies, organizational structures, publishing entities) in the academic and scientific realm. This attitude is employed to captivate diverse audiences with the aim of achieving professional benefits (e.g., securing funding, recognition, visibility, advancing careers), all in alignment with a neoliberal understanding of science. Further studies should further investigate the ramifications of these attitudes within different facets of the scientific community, particularly with a focus on their impact on scientific ethics.

**Limitations**

Future research should explore alternative data collection methods to mitigate potential social desirability bias in focus groups and validate these findings across diverse scientific fields and cultural contexts. Additionally, incorporating quantitative confirmatory research would bolster this line of inquiry.

**Acknowledgments**

This article is part of the research "Hype in scientific communication: credibility, reputation and well-being" funded by the Ministry of Education of Singapore (FY2021-FRC2-001).